\documentclass{SciPost}

\hypersetup{
    colorlinks,
    linkcolor={red!50!black},
    citecolor={blue!50!black},
    urlcolor={blue!80!black}
}
\usepackage[normalem]{ulem}
\usepackage[bitstream-charter]{mathdesign}
\DeclareSymbolFont{usualmathcal}{OMS}{cmsy}{m}{n}
\DeclareSymbolFontAlphabet{\mathcal}{usualmathcal}

\fancypagestyle{SPstyle}{
\fancyhf{}
\lhead{\colorbox{scipostblue}{\bf \color{white} ~SciPost Physics }}
\rhead{{\bf \color{scipostdeepblue} ~Submission }}

\fancyfoot[C]{\textbf{\thepage}}
}

\begin{document}

\pagestyle{SPstyle}

\begin{center}{\Large \textbf{\color{scipostdeepblue}{
Dynamical phase transition in generalized Dicke model with strongly interacting trapped Rydberg ions\\
}}}\end{center}

\begin{center}\textbf{
Manish Chaudhary\textsuperscript{1,$\star$}
Rejish Nath\textsuperscript{2}
and 
Weibin Li\textsuperscript{3} 
}\end{center}

\begin{center}
{\bf 1} Institut de Physique Nucléaire, Atomique et de Spectroscopie, CESAM, University of Liège, B-4000 Liège, Belgium
\\
{\bf 2} Department of Physics, Indian Institute of Science Education and Research, Pune 411008, India
 \\
{\bf 3} School of Physics and Astronomy, and Centre
 for the Mathematics and Theoretical Physics of Quantum Non-Equilibrium Systems, University of Nottingham, NG7 2RD, UK
\\[\baselineskip]
$\star$ \href{mailto:manish.phys123@gmail.com}{\small manish.phys123@gmail.com}\,\quad
\end{center}

\section*{\color{scipostdeepblue}{Abstract}}
\textbf{\boldmath{%
We study dynamical phase transitions in the generalized dissipative Dicke model in an array of trapped Rydberg ions, where their density-density interactions compete with the collective spin-phonon coupling, laser driving and dissipation. This setting offers a versatile approach to study equilibrium as well as non-equilibrium many-body phenomena, as parameters, such as the Ising interaction, laser-ion and spin-phonon coupling can be tuned. Through analyzing the mean-field phase diagram, we find a variety of distinct phases and the emergence of a tricritical point that are sensitively dependent of the interaction between Rydberg ions.  We then study the quantum dynamics for a finite system size and characterize parameter dependent dynamics using the spin average, entropy, and Loschmidt echo. Distinctive signatures of the dynamical phases, such as slow relaxation and metastability, arise near the phase transition. This analysis predicts rich quantum dynamics of the finite system that link to the non-equilibrium mean-field phases. Our study widens the exploration of collective and non-equilibrium phases in Dicke models, and reveals that the Rydberg ion interaction drastically affects the phase diagram and dynamics. 
}}

\vspace{\baselineskip}

\noindent\textcolor{white!90!black}{%
\fbox{\parbox{0.975\linewidth}{%
\textcolor{white!40!black}{\begin{tabular}{lr}%
  \begin{minipage}{0.6\textwidth}%
    {\small Copyright attribution to authors. \newline
    This work is a submission to SciPost Physics. \newline
    License information to appear upon publication. \newline
    Publication information to appear upon publication.}
  \end{minipage} & \begin{minipage}{0.4\textwidth}
    {\small Received Date \newline Accepted Date \newline Published Date}%
  \end{minipage}
\end{tabular}}
}}
}


\vspace{10pt}
\noindent\rule{\textwidth}{1pt}
\tableofcontents
\noindent\rule{\textwidth}{1pt}
\vspace{10pt}


\section{Introduction}
\label{sec:intro}
Dynamical phase transitions (DPTs) in open quantum systems have emerged as a central theme in non-equilibrium many-body physics~\cite{klinder2015dynamical,PhysRevLett.98.180601,PhysRevLett.110.135704,Heyl_2019,fazioSciPostSciPostPhys}, where steady states are determined by the competition between coherent dynamics and environmental coupling~\cite{RevModPhys.83.863,PhysRevLett.128.110603}. Unlike equilibrium phase transitions, which are governed by free-energy minimization, DPTs arise from the spectral properties of the Liouvillian superoperator that governs the density matrix evolution of the system ~\cite{PhysRevLett.110.135704,diehl2008quantum,PhysRevA.97.013825,heyl2018dynamical}. In particular, the closing of the Liouvillian gap—defined by the real part of the leading nonzero eigenvalue—signals critical slowing down and the onset of non-analytic behavior in steady-state observables~\cite{minganti2018spectral}. This framework has enabled the identification of rich phenomena such as bistability, metastability, and limit-cycle dynamics in driven–dissipative systems~\cite{PhysRevLett.105.015702,chaudhary2026,strogatz2024nonlinear,PhysRevA.85.043620,PhysRevE.96.053306}. Various theoretical and experimental studies have established the classification and signatures of such dynamical transitions in quantum optics ~\cite{tomadinReservoirEngineeringDynamical2012a,PhysRevA.85.063401,PhysRevA.75.013804,PhysRevLett.125.143602}, ultracold gases~\cite{greiner2002collapse,PhysRevLett.98.180601,PhysRevLett.107.140402} and many-body spin systems~\cite{PhysRevLett.110.135704,PhysRevA.97.013825,PhysRevLett.119.080501,karch2025probing}.

Trapped-ion systems provide a highly controllable setting for exploring these phenomena~\cite{PhysRevLett.119.080501,PhysRevLett.103.120502}, offering long coherence times, tunable interactions, and precise state preparation and readout capabilities ~\cite{RevModPhys.75.281}. When combined with Rydberg excitation to electronically high-lying states, trapped ions gain access to strong, long-range interactions~\cite{PhysRevLett.108.023003,PhysRevLett.123.153602,PhysRevA.101.052510,niederlander2023rydberg,zhang2020submicrosecond} that can be engineered via laser coupling to highly excited electronic states~\cite{muller2008trapped,PhysRevLett.132.223401,euchnerNonequilibriumExcitonDynamics2026}. This hybrid platform bridges the advantages of traditional trapped-ion quantum simulators~\cite{PhysRevLett.119.220501,RevModPhys.82.2313,shao2024rydberg} with the interaction strength that is typical of neutral Rydberg atom systems. Recent experimental advances have demonstrated the coherent control of Rydberg states in ion traps and the realization of interaction-driven many-body dynamics~\cite{PhysRevA.100.022513,photonics11121135}. These developments open a route to studying dissipative quantum dynamics, where both phonon-mediated couplings and Rydberg-induced interactions play significant roles ~\cite{mokhberi2020trapped,PhysRevLett.125.133602}.

In this context, driven–dissipative Dicke-type models implemented with trapped Rydberg ions provide a natural framework for investigating non equilibrium DPTs and their dynamic signatures~\cite{PhysRevA.100.022513,mokhberi2020trapped}. The interplay between collective spin and phonon coupling, spontaneous emission, and interaction-induced nonlinearities can give rise to coexistence regions~\cite{PhysRevA.100.022513,PhysRevA.85.013817}, intermittency, and tricritical behavior~\cite{PhysRevLett.120.183603} in the phase diagram of the system. Moreover, the Liouvillian spectrum offers direct insight into relaxation timescales~\cite{minganti2018spectral} and the emergence of criticality, whereas quantities such as the Loschmidt echo and subsystem entropies reveal the role of coherence and information scrambling in open-system dynamics~\cite{PhysRevLett.125.143602,karch2025probing}. In this work, we explore these aspects by analyzing the steady-state phases and dynamical evolution across different regimes of dissipation and interaction strength, highlighting how Rydberg interactions enhance dissipative stabilization and modify the nature of the underlying phase transitions.

Starting from a Dicke-type collective spin model, we analyze how dissipation and Rydberg-mediated interactions reshape quantum evolution, information scrambling, and relaxation toward steady states. Previous works have mainly focused mean field dynamics in thermodynamic limit~\cite{PhysRevLett.112.023603}, however in this work, we also analyze system dynamics with finite system size. By examining the spin expectation values, state populations, Loschmidt echo, and von Neumann entropy of the ion and phonon with different system sizes, we demonstrate a transition from coherent oscillatory dynamics in the closed system to dissipative behavior characterized by suppressed revivals and enhanced equilibration~\cite{strogatz2024nonlinear}.
A mean-field phase diagram reveals a tricritical behavior separating distinct dynamical regimes, indicating a competition between collective coupling, dissipation strength, and interaction-induced nonlinearities. Analysis of the Liouvillian spectrum shows that Rydberg interactions enlarge the effective dissipative gap and suppress long-lived oscillatory modes~\cite{PhysRevLett.105.015702,strogatz2024nonlinear}, leading to faster convergence to the steady state and reduced critical slowing down compared to purely dissipative dynamics. The phase transition is marked by the transient cusp singularity in the probability distribution
of the wavefunction occurring at the critical times and is the result of competing dynamic behaviors within the system.

The remainder of this paper is organized as follows: in Sec.~\ref{sec2}, we first introduce the system and physical Hamiltonian that we are dealing with. In Sec.~\ref{sec3}, we perform a mean field analysis and investigate the dissipative phase diagram resulting from the interplay among spin-phonon coupling, atomic dissipation, and Rydberg interactions.  We also examine the nonequilibrium dynamics following sudden quenches of the control parameters in the time evolution of the system. In Sec.~\ref{sec4}, we present exact numerical simulations of the density matrix dynamics and characterize the dynamical phase transitions by calculating various physical observables, including the collective spin expectation values, ion state populations, von Neumann entropy, Liouvillian spectrum, and Loschmidt echo. We also study the effect of the system size on the dynamics. 
In Sec.~\ref{sec5}, we summarize the main findings and conclude with future directions of the work.

\section{Physical system and model Hamiltonian}
\label{sec2}

\begin{figure}[t]%
\includegraphics[width=\linewidth]{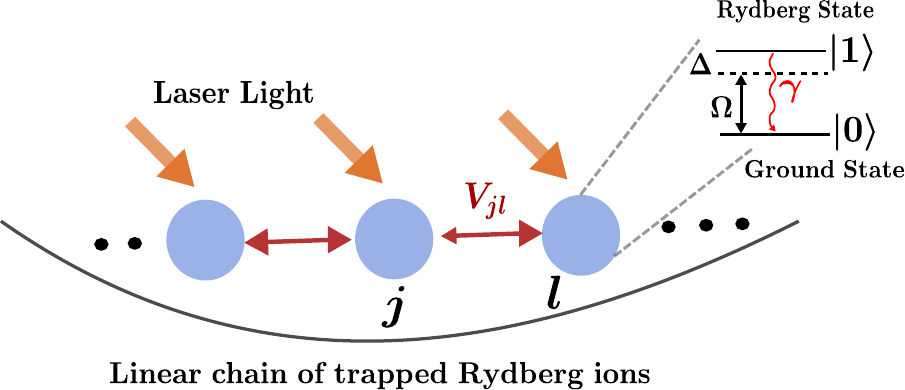}
\caption{Schematic illustration of the system of trapped Rydberg ions: A linear chain of $N$ trapped Rydberg ions is modeled as an effective two-level system with ground state $|0\rangle$ and excited Rydberg state $|1\rangle$.  A standing-wave laser couples the spin states to the center-of-mass (COM) vibrational mode at a frequency $\omega$. In the side panel, we represent the energy level diagram for the internal states which are coherently driven by a laser with Rabi frequency $\Omega$ and detuning $\Delta$. The state $|1\rangle$ decays to the ground state $|0\rangle$ at a spontaneous decay rate $\gamma$.
}
\label{fig1}%
\end{figure}

In this section, we discuss the physical system of a linear string of $N$ trapped Rydberg ions with two energy levels~\cite{PhysRevX.7.021038}. The lowest ground and excited Rydberg spin levels are denoted by $| 0 \rangle$  and $|1\rangle$, respectively, which are coupled by a laser with Rabi frequency $\Omega$ and detuning $\Delta$. Moreover, the internal spin states are coupled with the external vibrational modes (phonons) of the ion string~\cite{jamesQuantumDynamicsCold1998}, defined by a bosonic annihilation operator, $a$, with center of mass (COM) mode frequency $\omega$, via a standing wave laser field mediating a state-dependent spin-phonon coupling~\cite{PhysRevA.100.022513,muller2008trapped}. A schematic representation of the physical system is shown in Fig.~\ref{fig1}. The excited state decays to the lower ground state with a dissipation rate $\gamma$ such that dynamics of the density matrix for the system  is governed by the Master equation~\cite{schlosshauer2007decoherence,PhysRevB.102.115109,siebererUniversalityDrivenOpen2025}
\begin{align}
    \frac{\partial \rho}{\partial t} = -i [H, \rho] + \mathcal{L}(\rho),
    \label{eq:Master eq}
\end{align}
where $H$ represents the system Hamiltonian given by 
\begin{align}
    H = \Omega \sum_{j=1}^N \sigma^x_{j} + \Delta \sum_{j=1}^N \sigma^z_{j}  + \sum_{j,l\neq j} V_{jl} \sigma^z_{j} \sigma^z_{l} + g \sum_{j=1}^N \sigma^z_{j} (a + a^\dagger) + \omega  a^{\dagger} a,
     \label{eq:Ham}
\end{align}
such that $\sigma^\alpha_{j}$ represents the Pauli matrix for the $j^{\text{th}}$ ion and $\alpha \in \{x,y,z\}$. The occupation number for the internal states is defined by the operator $n_j$. The dipole-dipole interaction strength between $j^{\text{th}}$ and $l^{\text{th}}$ ion  is defined by $V_{jl} \sim \frac{1}{|\textbf{r}_j - \textbf{r}_l|^3}$ in~\cite{Gallagher_1994}, while $g$ represents spin-phonon coupling. 

The Liouvillian for the Markovian dissipative process in (\ref{eq:Master eq}) is defined by
\begin{align}
    \mathcal{L}(\rho)  = \frac{\gamma}{2} \sum_{j=1}^N \left( 2 \sigma_j^- \rho \sigma_j^+ -  \{ \sigma_j^+ \sigma_j^-, \rho \} \right),
    \label{eq:liouv}
 \end{align}
with the spin ladder operator $\sigma_j^\pm = \frac{\sigma_j^x \pm i \sigma_j^y}{2}$.
With $V_{jl} = 0$, Eq. (\ref{eq:Ham}) represents the usual Dicke model that has been studied in detail in~\cite{PhysRevA.75.013804,PhysRevA.71.053804}.


\section{Dissipative Phase diagram and quenching dynamics}
\label{sec3}
We first review the key features of the model as described by Eq.~\eqref{eq:Master eq} and understand the importance of each interaction term in the Hamiltonian~\eqref{eq:Ham}.  This is achieved through a mean-field (MF) analysis and we calculate the time-averaged dynamics of the spin expectation values. Furthermore, we study the non equilibrium response of the system to sudden quenches of the control parameters~\cite{PhysRevLett.119.080501}, revealing their influence on the transient and long-time dynamical behavior.

\subsection{Mean field analysis}
We begin by analyzing the steady state properties of the dissipative model governed by Eq.~\eqref{eq:Master eq} in the MF limit. Within this approximation, the COM phonon mode is represented by the classical variables corresponding to its average displacement $X$ and average momentum $P$  as,
\begin{align}
    X & = \frac{\langle a \rangle + \langle a^\dagger \rangle}{2} ,\nonumber \\
    P & = \frac{\langle a \rangle - \langle a^\dagger \rangle}{2i},
    \label{eq:MF_phonons}
\end{align}
while the average spin value of the ions is defined as,
\begin{align}
    S_{x,y,z} = \frac{1}{N}\sum_j \langle \sigma^{x,y,z}_j \rangle.
    \label{eq:MF_spins}
\end{align}
Any product of spin and bosonic operator is approximated as $\langle \sigma^{x,y,z}_j a \rangle \approx \langle \sigma^{x,y,z}_j \rangle \langle a \rangle$ and $\langle \sigma^{x,y,z}_j \sigma^{x,y,z}_l \rangle \approx \langle \sigma^{x,y,z}_j \rangle \langle \sigma^{x,y,z}_l \rangle$, under this simplification we obtain the following set of mean field equations~\cite{kolomietz2020mean,sakthivadivel2022magnetisation},
\begin{align}
    \dot{X} & = \omega P,  \nonumber \\
    \dot{P} & = gNS_z - \omega X, \nonumber \\
    \dot{S}_x & = -f(X,P)S_y -2 \mathcal{V} S_z S_y - \frac{\gamma}{2} S_x ,\nonumber \\
    \dot{S}_y & = -f(X,P)S_x -2 \mathcal{V} S_z S_x -2\Omega S_z -\frac{\gamma}{2} S_y ,   \nonumber \\
    \dot{S}_z & = 2\Omega S_y - \gamma (1+S_z),
    \label{eq:meanfield}
\end{align} 
where the MF interaction potential is expressed as $\mathcal{V}=\frac{1}{2N}\sum_{j \neq l} V_{jl}$, and $f(X,P)\\= 2(\Delta+2gX )$. We choose $\omega = N \Omega$ to make the phase boundaries independent of $N$.

\begin{figure}[t]%
\includegraphics[width=\linewidth]{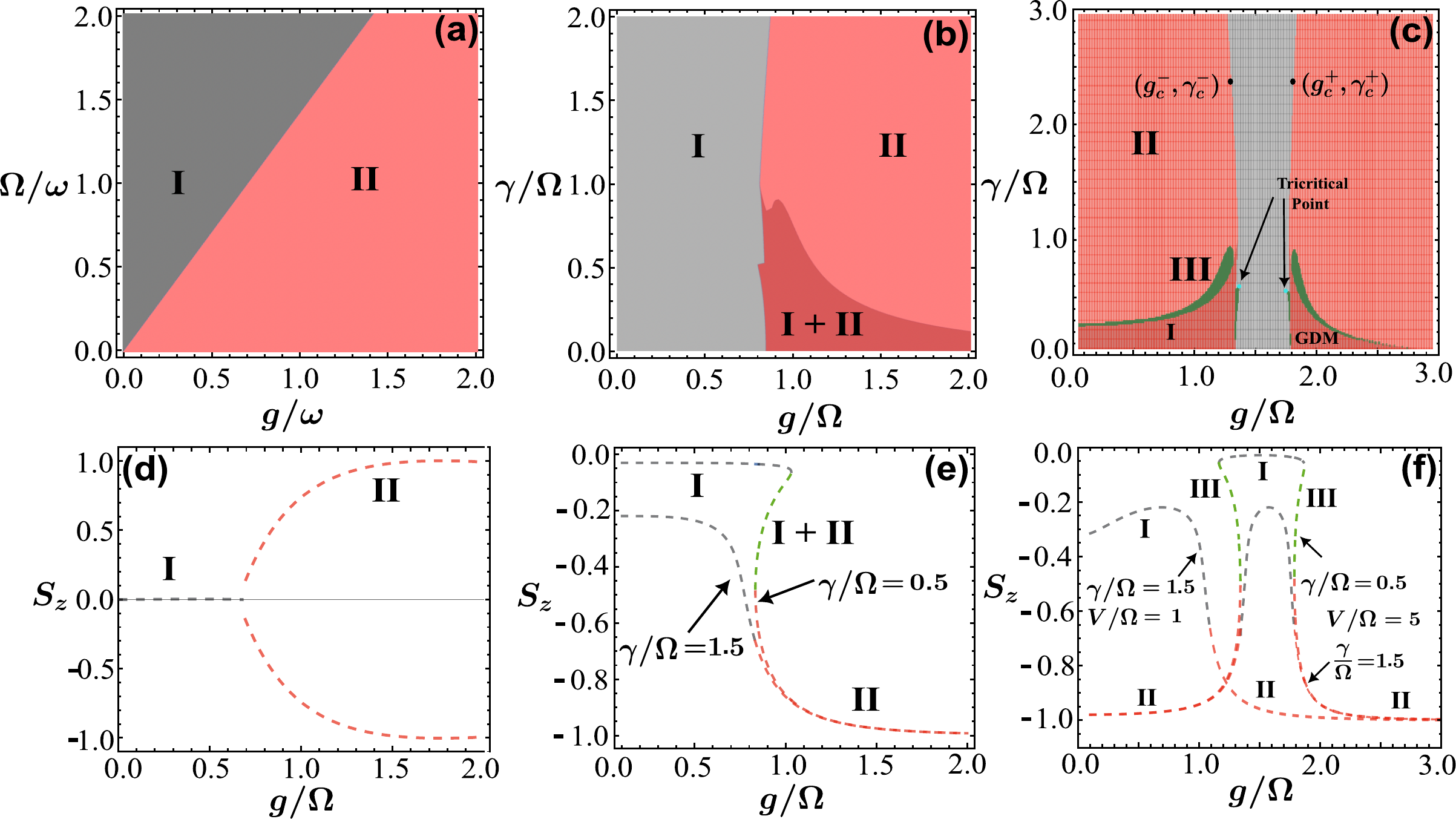}
\caption{Dynamical phase diagram for different controlling parameters \eqref{eq:Master eq} in the dissipative model:  
The MF phase diagram is obtained from the fixed-point analysis of Eq. \eqref{eq:meanfield} (a) as a function of the spin phonon coupling $g/\omega$ and Rabi frequency $\Omega/\omega$ with no dissipation  and no Rydberg ion interactions $\mathcal{V} = 0$ (Dicke Model); (b) as a function of normalized parameters $g/\Omega$ and $\gamma/\Omega$ with ion decay \eqref{eq:liouv} (dissipative Dicke model); (c) as a function of normalized parameters $g/\Omega$ and $\gamma/\Omega$ with ion interactions $V_{jl}\neq 0$ (generalized dissipative Dicke model); (d)-(e) Corresponding variation of MF spin expectation value $S_z$ as a function of $g/\Omega$ for different cases similar to (a)-(c) respectively. For cases (b)-(f) we choose $\omega=\Omega N$ and $\Delta=0$. The phase diagram is classified into different stable regions. Distinct phases are marked as Roman letters.}
\label{fig2}%
\end{figure}

The steady state solutions are obtained from (\ref{eq:meanfield}) by setting first derivatives to zero. The resulting fixed points are analyzed through a linear stability analysis, and the stable solutions are identified as distinct dynamical phases, giving rise to the MF phase diagram as shown in Fig.~\ref{fig2}. 
We briefly summarize the dynamics and nature of the stable solutions associated with each interaction term in the Hamiltonian model~\eqref{eq:Ham} which has been analyzed in detail in~\cite{PhysRevA.100.022513}. 

\subsection{Closed and dissipative Dicke model}
By setting the Rydberg interaction and dissipation to zero, as $V_{ij} = 0,\gamma=0$ in Eqs.~\eqref{eq:Ham}-\eqref{eq:liouv} respectively, the model reduces to the conventional closed Dicke Hamiltonian~\cite{kirton2019introduction,PhysRevA.69.053804}.  This model has been extensively studied as a fundamental framework for describing collective light-matter interactions, describing an ensemble of two-level systems coupled to a single bosonic mode in the context of equilibrium quantum phase transitions~\cite{kirton2019introduction}.  This model marks a quantum phase transition from a normal phase (Phase I) to a superradiant phase (Phase II), as shown in the phase diagram in Fig. \ref{fig2}(a). 
Two stable solutions are obtained by solving MF equations~\eqref{eq:meanfield} that characterize the spin average value along $z$-axis as
\begin{align}
    S_z = 0, \hspace{1cm}  S_z= \pm \frac{\sqrt{4 g^4-\Omega ^4}}{2 g^2} 
    \label{MF_expec_values}
\end{align}
such that Phase I is characterized by vanishing collective bosonic displacement and spin polarization value, while Phase II occurs when both the bosonic field and collective spin acquire finite expectation values as in Eq.~\eqref{MF_expec_values}.
Figure~\ref{fig2}(d) shows the variation of the spin expectation value $S_z$ as a function of the  scaled spin-phonon coupling strength $g/\Omega$ which shows a change in the population imbalance from a negligible value in Phase I to a macroscopic value in Phase II. The corresponding critical coupling is given by
\begin{align}
 g_c=\sqrt{\frac{\omega \Omega}{2N}} .
\end{align}
A finite detuning~\cite{PhysRevA.69.053804} results in a crossover between two phases and smoothens out the phase boundary.

Considering the finite life time of the excited state ($\gamma \neq 0$), the system exhibits qualitatively different dynamical behavior in dissipative regime~\eqref{eq:liouv}, and MF analysis reveals distinct phases, as illustrated in Fig.~\ref{fig2}(b). The driven dissipative Dicke model~\cite{klinder2015dynamical} is governed by the competition between coherent spin-phonon coupling and excited state decay, giving rise to two characteristic steady-state phases: a bright phase (Phase I), characterized by a vanishing phonon displacement $(X=0)$ and dominated by the interplay between coherent driving and dissipation, and a dark phase (Phase II), where the spin-phonon coupling suppresses both driving and decay, resulting in a finite center-of-mass displacement $(X>0)$. Bright and dark phases are detected as high photon and low photon emission respectively in the fluorescence signal. In this case, the stability of steady state solutions infer that
\begin{align}
    g_c = \frac{\sqrt{\omega}(3\gamma ^2 + 16 \Omega ^2)^{1/4} }{2 \sqrt{2N}}  .
\end{align}
The stability analysis of the MF equations further predicts a coexistence region (Phase I $+$ II), where both bright and dark steady states are stable~\cite{PhysRevA.85.013817}. This bistability leads to a strong dependence  of dynamics on the initial state, and the dynamics also features intermittent switching between bright and dark dynamical trajectories. This coexistence leads to intermittent fluorescence in the ion emission signal~\cite{PhysRevLett.112.023603}.
The corresponding region is shown by the overlapping red and gray area in Fig.~\ref{fig2}(b).
The average spin values are plotted in Fig. \ref{fig2}(e) and we find that the coexistence phase vanishes at larger dissipation value of $\gamma/\Omega=1.5$.
A finite detuning~\cite{PhysRevLett.112.023603} doesnot destropy DPTs in this case (Ref.~\ref{fig2}(b)) and MF steady states remain qualitatively unchanged.


\subsection{Generalized dissipative Dicke model with Rydberg interactions} 
Figure \ref{fig2}(c) presents the dissipative phase diagram illustrating the influence of strong Rydberg interactions ($\mathcal{V} \neq 0$) on non equilibrium dynamics as defined by the model Hamiltonian~\eqref{eq:Ham}. The presence of inter-ion interactions introduces nonlinearities that substantially alter the steady state landscape. The Rydberg interaction adds a energy scale that competes with the spin-phonon coupling and dissipation generating multiple stable MF solutions. This gives rise to bistability, coexistence regions, and the emergence of new critical points~\cite{PhysRevA.100.022513}, these features are entirely absent in the simplest dissipative Dicke model~\cite{klinder2015dynamical}. Mathematically, the steady state spin polarization $S_z$ satisfies a \textit{cubic} polynomial equation
\begin{align}
\sum_{j=0}^3 a_j S_z^j = 0 ,   
\label{eq:Szsteady}
\end{align}
where the coefficients $a_j$ depend on the system parameters. We observe that critical points emerge in the parameter region as,
\begin{align}
    (g_c^{\pm}, \gamma_c^{\pm}) = \left( \sqrt{\frac{\omega \Omega}{2N}} \sqrt{\frac{\mathcal{V}}{\Omega}\pm\frac{\sqrt{27}}{4}},\Omega \right).
    \label{eq:critical points}
\end{align}
The emergence of a new critical point occurs for an interaction value $\mathcal{V}>\sqrt{27}\Omega/4$ indicating the onset of dynamics due to Rydberg ion interactions.

For strong spin-phonon coupling $g/\Omega$ and weak atomic decay $\gamma/\Omega$, one obtains two stable solutions for \eqref{eq:Szsteady} linked to the usual bright and dark phases (Phase I and Phase II), and the third solution is unstable as it gives the positive eigenvalue of Jacobian matrix for linearized equation~\cite{strogatz2024nonlinear}. This is also confirmed by the $S_z$ average plot in Fig. \ref{fig2}(f). This region is associated with the critical point ($g_c^{+}, \gamma_c^{+}$). Rydberg interactions stabilize a Generalized Dicke Model (GDM) phase, which is absent in the conventional dissipative Dicke model and exhibits finite spin polarization together with a nonzero phonon field at larger spin-phonon coupling. 

In contrast, the coexistence region associated with the second critical point ($g_c^{-}, \gamma_c^{-}$) arises solely due to finite inter ion interactions, and has properties fundamentally different from those of the bistable region described above.  Between these phases lies a phonon-lasing coexistence region (Phase III), where no stable fixed point exists and the system evolves into persistent limit-cycle oscillations, giving rise to intermittent fluorescence and self-sustained nonequilibrium dynamics. 
As illustrated in Fig.~\ref{fig2}(f), reducing the interaction strength gradually suppresses this interaction-induced coexistence region~\eqref{eq:critical points}, eventually eliminating the critical point ($g_c^{-}, \gamma_c^{-}$). The central grayish area corresponds to the phonon-lasing coexistence region with Phase I.

Furthermore, a small detuning $\Delta$ does not change the phase diagram significantly, however a larger value reshapes the nonequilibrium phase diagram by altering the stability of steady-state solutions~\cite{PhysRevA.100.022513}. In general, negative detuning ($\Delta<0$) suppresses the interaction-induced coexistence region, whereas positive detuning ($\Delta>0$) destabilizes the GDM coexistence region. The additional coexistence regimes~\cite{PhysRevB.100.125102} and a richer dynamical landscape can be introduced by presence of a state-dependent trapping potential. More details about the dynamics in this case can be found in Ref. ~\cite{PhysRevA.100.022513}.

One striking feature of this dissipative model is the presence of a tricritical point where all three phases coexist as seen in Fig.~\ref{fig2}(c). One can see that the transition lines meet at 
\begin{align}
    (g/\Omega, \gamma/\Omega) & = (1.36,0.57),\nonumber \\
    (g/\Omega, \gamma/\Omega) & = (1.77,0.52).
\end{align}
A related dissipative phase transition has been demonstrated in Ref.~\cite{PhysRevLett.120.183603}, where a generalized Dicke Hamiltonian is considered in which two independent quadratures of the cavity field couple to different components of the collective spin. In contrast, this study considers a trapped Rydberg-ion system with state-dependent spin-phonon coupling and Rydberg-mediated interactions, providing an alternative physical platform for investigating dissipative phase transitions and interaction-induced nonequilibrium phenomena.

\begin{figure}[t]%
\includegraphics[width=\linewidth]{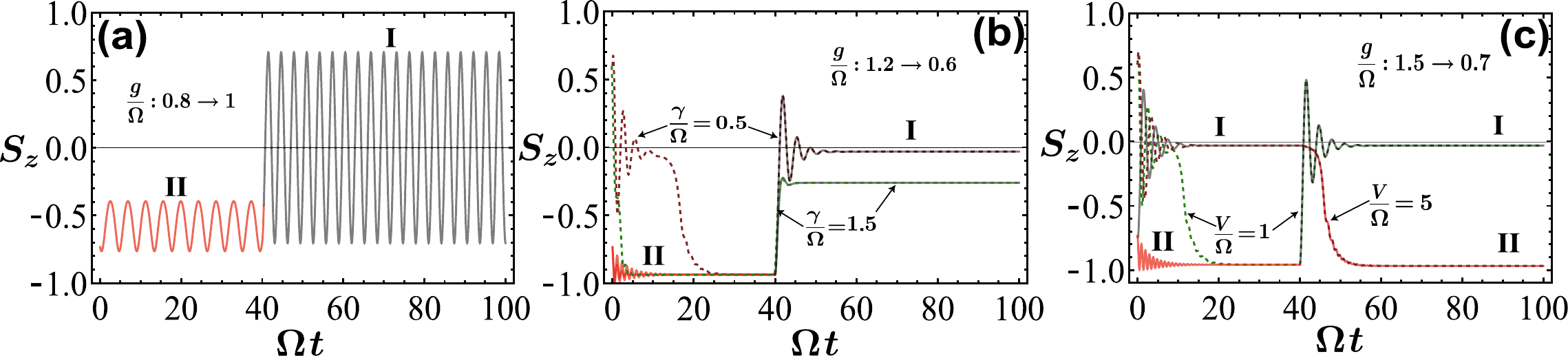}
\caption{Effect of quenching on the system dynamics for (a) simplest Dicke model with no Rydberg ion interactions $\mathcal{V} = 0$, (b) dissipative Dicke model (\ref{eq:liouv}) for two different $\gamma/\Omega=0.5,1.5$ and (c) generalized Dicke model with ion interactions $V_{jl}\neq 0$ at $\gamma/\Omega=0.5$. For all cases we choose $N=30$ and $\omega=N \Omega$. The distinct phases are marked with Roman letters. The different initial conditions are used for our calculation: ($S_x^0=0,S_y^0=0.4,S_z^0=0.3$) represented as dotted line and ($S_x^0=-0.67,S_y^0=-0.08,S_z^0=-0.73$) represented as solid line.
}
\label{fig3}%
\end{figure}

\subsection{Quenching dynamics}
In this section, we discuss the effect of quenching~\cite{PhysRevLett.125.040602,PhysRevB.111.184319,ali2024quantum} on the time dynamics of the system when the control parameter suddenly changes around a critical point in each case. 

To investigate nonequilibrium dynamics, we define quench time $t_q$ as the instant at which the value of the controlling parameter is allowed to change abruptly as,
\begin{align}
g = \left\{
\begin{array}{cc}
g_-  & \hspace{1cm} t < t_q \\
g_+  & \hspace{1cm} t\ge t_q .
\end{array}
\right.  
\end{align}
which is achieved by changing the intensity or frequency of the laser field that mediates the state-dependent force acting on the trapped ion system~\cite{kulosa2023systematic,PhysRevA.51.R2699}.

Fig. \ref{fig3}(a) captures the effect of quenching the spin-phonon coupling parameter in the simplest Dicke model (Fig. \ref{fig2}(a)) for a particular initial state. We compute the time evolution of the spin operator $S_z$ and change the coupling parameter at the quench time $t_q$. Before quenching $t<t_q$,the system resides in Phase II, characterized by a finite macroscopic value of $S_z$. After quenching $t>t_q$, the dynamics transition to Phase I, where the time-averaged value of $S_z$ is zero, as it oscillates symmetrically between positive and negative values. The persistent oscillations observed after the quench that originate from the coherent, non-dissipative nature of the closed Dicke dynamics as seen in Fig.~\ref{fig2}(a).

Fig. \ref{fig3}(b) illustrates the effect of sudden quench in the dissipative Dicke model (Fig. \ref{fig2}(b)) for two different initial states. Owing to the presence of dissipation, the coherent oscillations observed in the closed Dicke model are suppressed, and the system relaxes toward well-defined steady states both before and after the quench. For the dissipation parameter $\gamma/\Omega=0.5$, the system initially evolves to the dark phase (Phase II), characterized by a steady state with
\begin{align}
     S_z=-1.
\end{align}

At quench time $t=t_q$, the spin-phonon coupling is switched to a regime that causes the system to relax to a new steady state with $S_z=0$ (Phase I), though the transition is not abrupt because of the bistable regime. This is true for any initial state as shown in Fig. \ref{fig3}(b).
We also study the effect of varying the dissipative parameter on the quench dynamics. The dynamics is comparatively similar to the previous case, however for larger dissipation ($\gamma/\Omega=1.5$), the transition to the steady state is sharp, remarking the stronger influence of dissipation in the driven open system dynamics. 
 
Figure~\ref{fig3}(c) illustrates the quench dynamics of the interacting dissipative model (Figs.~\ref{fig2}(c)) in the presence of finite Rydberg-mediated ion-ion interactions. We investigate the nonequilibrium response of the system by suddenly quenching the spin-phonon coupling strength. For a strong interaction $V/\Omega=5$, the system initially resides in the bright phase (Phase I), characterized by $S_z=0$ before $t<t_q$. However, the macroscopic spin state with $S_z=-1$ is obtained after the quench time $t>t_q$. One can see that the transition between the two phases is not abrupt but occurs over an extended transient timescale which originates from the interaction-induced coexistence region associated with the critical point ($g_c^{-}, \gamma_c^{-}$). Reducing the interaction strength to $V/\Omega=1$, qualitatively modifies the quench dynamics. In this regime, the system undergoes the reverse transition, evolving from the dark phase (Phase II) to the bright phase (Phase I) following the quench. This is attributed to the disappearance of the interaction-induced coexistence region which weakens the metastability.


\section{Finite system size dynamics}
\label{sec4}
In the previous section, we investigated the fixed points, nonequilibrium dynamics, and phase structure of the generalized Dicke model within the MF approximation. We now go beyond this limit and examine the finite-size dynamics by considering systems with varying numbers of ions, thereby elucidating the role of the quantum fluctuations.
We calculate the exact time evolution of the system's density matrix (\ref{eq:Master eq}) by using the QuTiP Python package \cite{PhysRevA.98.063815}. 
To characterize the dissipative dynamics and identify signatures of dynamical phase transitions, we introduce the general definitions of an important physical quantity, the Loschmidt Echo~\cite{karch2025probing,Goussev:2012} as a measure of the fidelity of the time-evolving quantum state and use it to probe the nonequilibrium response of the system. We further analyze the collective spin expectation spin, Von Neumann entropy and population of ion levels. 

\subsection{Spin average  and population of ion levels}
We calculate the time evolution of the exact spin expectation values from the Master equation \eqref{eq:Master eq} as 
\begin{align}
     S_\alpha (t)  = \text{Tr}[S_\alpha \rho(t)],
    \label{spin_exp_finite}
\end{align}
where $\rho(t)$ is the density matrix of the system, which is obtained by solving the master equation \eqref{eq:Master eq}  numerically.
The initial state of the system is choosen as,
\begin{align}
    |\psi_0\rangle = |\phi_0\rangle\otimes |n\rangle ,
\end{align}
such that the corresponding initial density matrix is
\begin{align}
   \rho(0) =  |\psi_0\rangle \langle \psi_0|
\end{align}
where the ions are initialized in the ground state $|\phi_0\rangle = |0,0\dots 0\rangle$, and $|n\rangle$ represents the Fock state for the common mode of phonons.

\begin{figure}[t]%
\includegraphics[width=\linewidth]{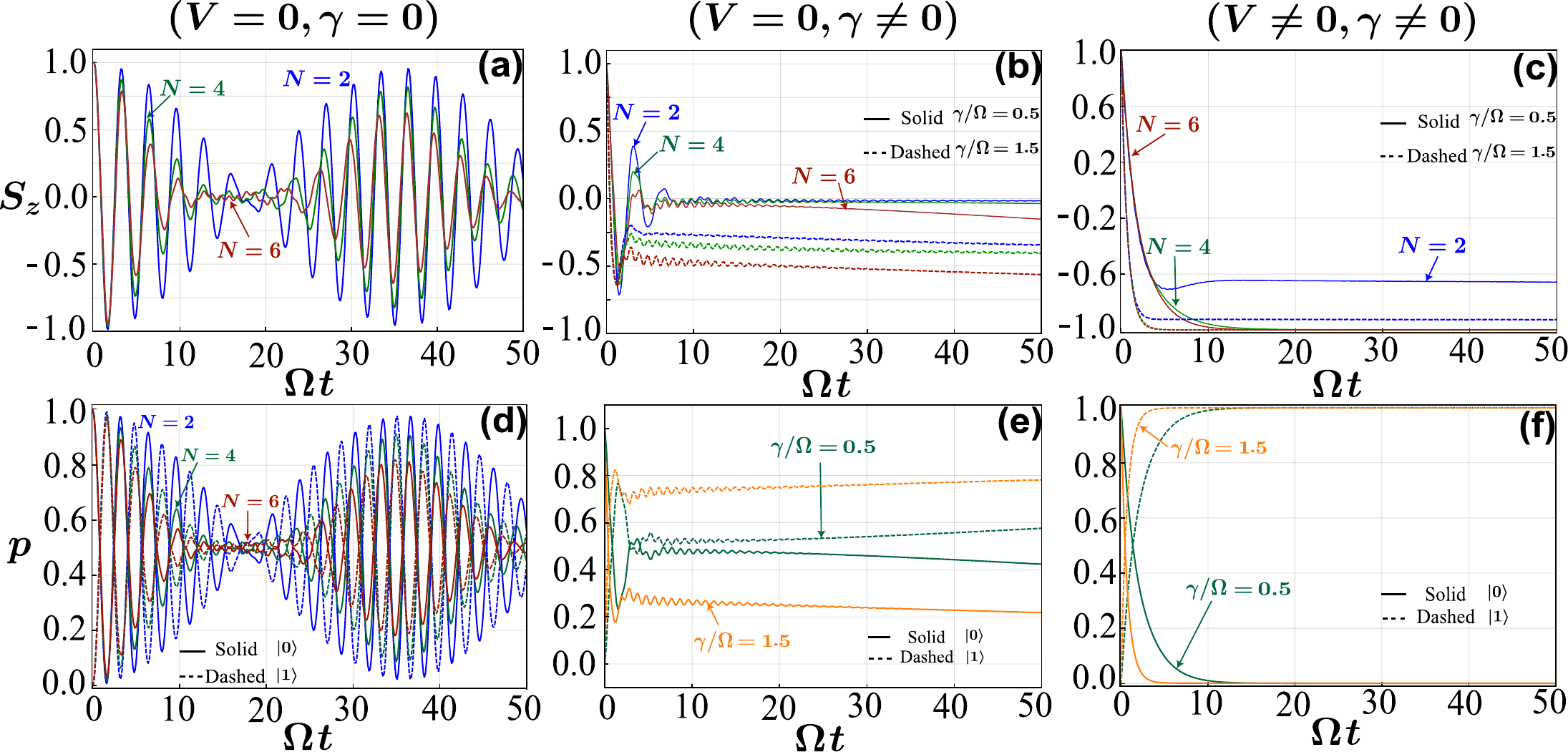}
\caption{Dynamics of dissipative model \eqref{eq:liouv}:  (a)-(c) Exact time evolution of the density matrix, spin expectation values (\ref{spin_exp_finite}); (d)-(f) probability of occupancy of the ground and excited states ($N=6$ used in (e),(f)) as a function of time $t$ with no dissipation ($V=0,\gamma=0$), with dissipation ($V=0,\gamma\neq0$) and with Rydberg interactions ($V\neq0,\gamma\neq0$) respectively for various system sizes. We choose $g/\Omega=0.5$  for all cases and $\mathcal{V}/\Omega=5$.
}
\label{fig4}%
\end{figure}

Fig \ref{fig4} captures the time variation of the $S_z$ spin-averaged value and population of the ground and excited states of the ions in various interaction and dissipative regimes. Fig. \ref{fig4}(a) considers the simplest Dicke model where we see Rabi's oscillations between the ground $|0\rangle$ and excited state $|1\rangle$ near the fixed value of critical spin-phonon coupling $g_c$. This occurs as the dynamics is dominated by a nearly resonant spin-phonon interaction, resulting in almost periodic population oscillations. We also plot the spin averaged values with varying ion number $N$ in a chain and we notice that the amplitude of the Rabi's oscillations decreases, this amplitude decrement is expected as the Hilbert space is larger with increasing $N$. Similarly, as shown in Fig. \ref{fig4}(d) that the population of the ions oscillates between the ground (solid line) and excited state (dashed line).  

However, the dynamics of the system is greatly affected in the presence of dissipation when the excited state is decayed to the ground state near the critical point value $g_c$. In Fig. \ref{fig4}(b), we plot the time variation of $S_z$ values with two different decay parameters $\gamma/\Omega =0.5,1.5$. For the small dissipation, Rabi's oscillations are damped such that the state is a steady state at longer times near $S_z=0$, this is true for any initial state and is generally not dependent on the ion number $N$. In contrast, with a relatively larger dissipation parameter, one expects a sharp dampening of the amplitude associated with Rabi oscillations, and the state of the system is prepared in different steady states, as shown in Fig. \ref{fig4}(b) with 
\begin{align}
    S_z<0
\end{align}
This is also supported by the time variation in the ion populations between different levels, as shown in Fig. \ref{fig4}(e). We note that the ion is prepared approximately in a superposition state with a probability $0.5$ for the occupancy in $|0\rangle$ and $|1\rangle$ states simultaneously with low dissipative decay (Phase I); however, the larger dissipation favors the majority of the ions to be pushed towards the excited state, marking Phase II, as shown in Fig. \ref{fig2}(e).

In the presence of Rydberg ion interactions ($V/\Omega$), qualitative and quantitative changes in the dynamics are observed. The interaction stabilizes a strongly polarized steady state, where $S_z$ rapidly saturates and remains nearly time-independent, as shown in Fig. \ref{fig4}(c). Compared with the purely dissipative case, interactions suppress residual oscillations and induce faster convergence toward ordered configurations due to the dephasing effects from the Rydberg-Rydberg interactions ~\cite{raitzsch2009investigation}. The probability evolution shown in  Fig. \ref{fig4}(f) reveals interaction-induced state selection: depending on the dissipation strength, the system is driven toward either the ground-state or excited-state dominated population. Increasing the system size enhances this stabilization, as the collective interactions provide synchronization among the spins and reduce the fluctuations. We can observe the crossover from coherent oscillatory dynamics (closed Dicke limit) to dissipative relaxation and finally to interaction-stabilized steady states, where many-body correlations and dissipation jointly determine the long-time behavior.

\subsection{Liouvillian spectrum}
Studying the Liouvillian spectrum~\cite{minganti2018spectral,campaioliQuantumMasterEquations2024} is crucial for understanding the dynamics of the system. Liouvillian superoperator satisfies the following eigenvalue equation 
\begin{align}
    \mathcal{L} \rho = \lambda \rho.
 \end{align}
The steady state corresponds to the eigenvalue $\lambda_0=0$, while the next lower eigenvalue $\lambda_1$ sets the slowest relaxation rate of the open system towards its steady state, also known as the Liouvillian gap. 

\begin{figure}[t]%
\includegraphics[width=\linewidth]{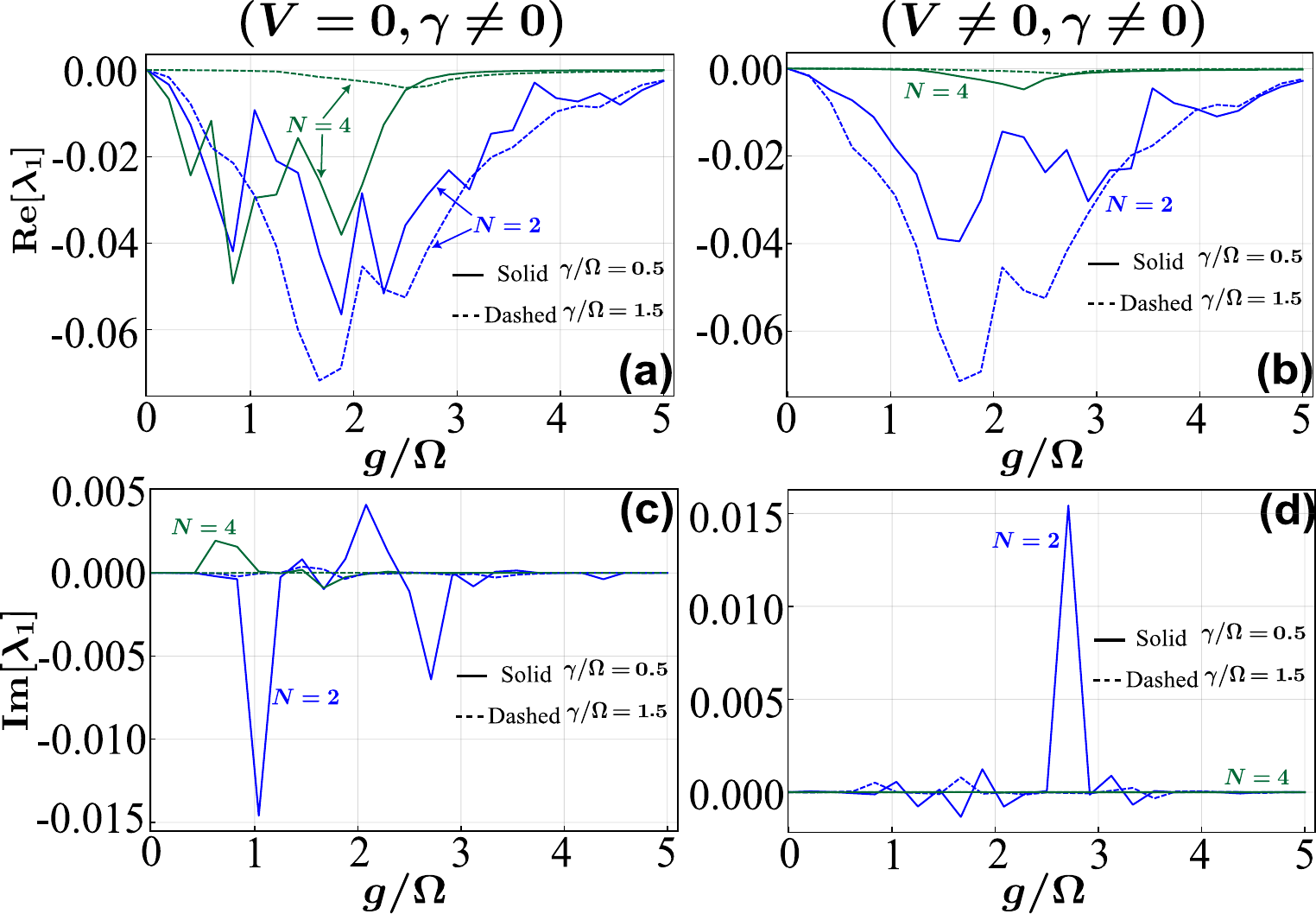}
\caption{Liouvillian spectrum: (a),(b) real part of the lowest nonzero Liouvillian eigenvalue $\text{Re}[\lambda_1]$; (c),(d) imaginary part of the lowest nonzero Liouvillian eigenvalue $\text{Im}[\lambda_1]$ as a function of the normalized spin-phonon coupling $g/\Omega$ in the absence and presence of Rydberg interactions respectively for different system size $N$ and dissipation strengths $\gamma/\Omega$. The solid and dashed lines denote the dynamics for dissipation strengths $\gamma/\Omega=0.5$ and $\gamma/\Omega=1.5$ respectively. For (b),(d) cases we choose $\mathcal{V}/\Omega=5$.
}
\label{fig5}%
\end{figure}

Fig. \ref{fig5}(a) plots the real part of the first non-zero Liouvillian eigenvalue $\text{Re}[\lambda_1]$ as a function of the normalized spin phonon coupling $g/\Omega$ for various system sizes $N$ and dissipation strength $\gamma/\Omega$. For small $g/\Omega<<1$, the value is close to zero, reflecting a weak spin-phonon connection and slow relaxation. As $g/\Omega$ increases, the value of $\text{Re}[\lambda_1]$ becomes more negative, indicating a faster decay due to the dissipative channels. This effect is stronger for larger $\gamma/\Omega$, which uniformly pushes the spectrum downward. We also observe the effect of varying $N$, for small $N$, the gap exhibits pronounced minima and remains finite over a broad range of $g/\Omega$, while for larger $N$, the minimum becomes shallower and the curve approaches zero again at intermediate and large $g/\Omega$. We observe that increasing $N$ suppresses the relaxation rates and drives the system towards a thermodynamic regime in which the Liouvillian gap closes in the large $N$ limit.

The imaginary part of the first non-zero Liouvillian eigenvalue $\text{Im}[\lambda_1]$ encodes the oscillation frequency of the slowest dynamic mode and indicates the presence of oscillatory dynamics. We plot it as a function of $g/\Omega$ as shown in Fig. \ref{fig5}(c). For small system sizes, the nonzero values of $\text{Im}[\lambda_1]$ over a finite range of coupling $g/\Omega$ indicate that the dominant relaxation mode is oscillatory, and hence, coherent spin–phonon exchange survives despite dissipation. An increase in the dissipation strength $\gamma/\Omega$ suppresses the oscillations. We observe that for larger $N$, $\text{Im}[\lambda_1]$ is strongly reduced and rapidly approaches zero for all couplings as $ N$ increases. This vanishing value of the imaginary part reflects the regime in which the fluctuations and coherent oscillations are overdamped.

When Rydberg interactions are introduced ($V\neq 0$), the Liouvillian spectrum becomes noticeably more regular and stable, as shown in Fig. \ref{fig5}(b),(d). The value $\text{Re}[\lambda_1] \to 0$ more smoothly as $g/\Omega$ increases, and the deep minima seen in the non-interacting dissipative case are reduced. The imaginary part $\text{Im}[\lambda_1]$ is also strongly suppressed, indicating that the coherent oscillatory channels are efficiently damped. Thus, Rydberg interactions generate dynamics with a larger effective dissipative gap and a more stable convergence to the steady state, as also seen in the spin average and population imbalance dynamics (Fig.~\ref{fig4}(c),(f)).

\subsection{Loschmidt Echo}
The Loschmidt echo is defined as the absolute value of the quantum mechanical overlap of the time-evolved state with the initial state~\cite{gorin2006dynamics,Goussev:2012,zhouSciPostSciPostPhys} as 
\begin{align}
    L = |\langle \psi (t)|\psi(0)\rangle|.
\end{align}
For dissipative dynamics, we define it in a similar way using density matrix,
\begin{align}
     L(t) = -\frac{1}{N}\ln{\text{Tr}[\rho(0)\rho(t)]} ,
     \label{eq:lecho}
\end{align}
It is an important quantity that shows a non-analytic behavior with time that indicates dynamical phase transitions~\cite{schusterOperatorGrowthOpen2023a,karch2025probing}. It measures the sensitivity of the time evolution of the system to its dynamics and can be interpreted as the revival probability of the initial state of the system.  
\begin{figure}[t]%
\includegraphics[width=\linewidth]{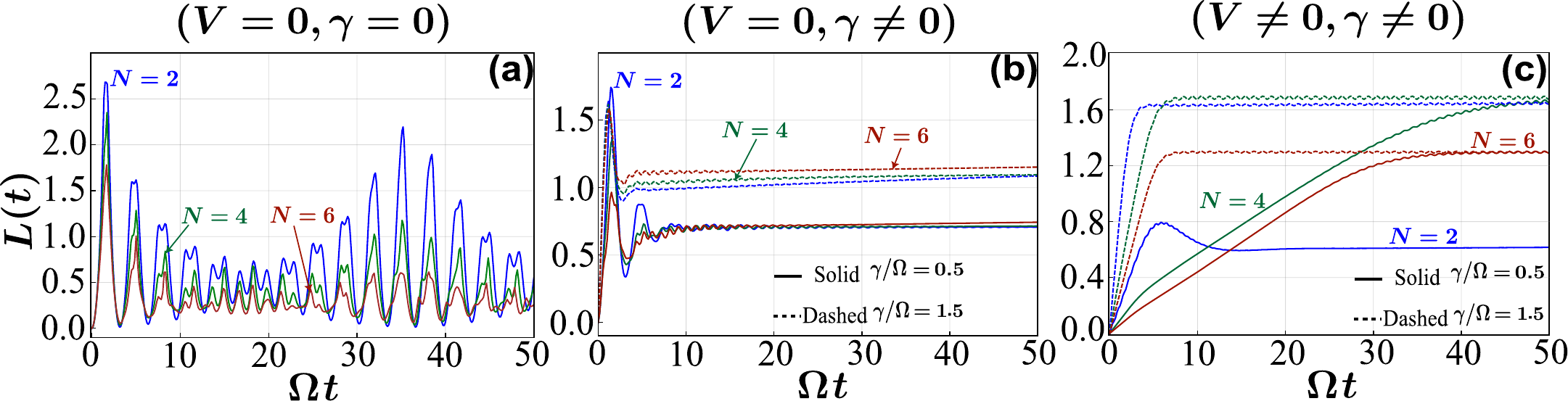}
\caption{Time evolution of the Loschmidt echo~\eqref{eq:lecho} for different system sizes $N$: (a) Closed Dicke dynamics ($V=0,\gamma=0$) showing coherent oscillations and revivals; (b) dissipative Dicke dynamics ($V=0,\gamma \neq 0$) leading to damped oscillations and a steady-state plateau; and (c) dissipative dynamics with Rydberg interactions ($V \neq 0,\gamma \neq 0$) exhibiting interaction-induced relaxation toward a stable steady value $L_s$. The solid and dashed lines denote the dynamics for dissipation strengths $\gamma/\Omega=0.5$ and $\gamma/\Omega=1.5$ respectively.  We choose $g/\Omega=0.5$  for all cases and $\mathcal{V}/\Omega=5$.
}
\label{fig6}%
\end{figure}

Fig. \ref{fig6}(a) shows the time variation of Loschmidt echo with varying system sizes for the simplest Dicke model. We observe that the local peaks appearing at quasi-regular time intervals indicate partial revivals, where the system’s wave function overlaps with the initial state. These revivals are characteristic of coherent dynamics induced by spin–phonon interactions which set the recurrence times.  
For larger system sizes, the amplitude of Loschmidt echo vanishes due to the bigger Hilbert space, which implies a faster loss of memory of the initial state.

Fig. \ref{fig6}(b) shows the time variation of Loschmidt echo in the presence of dissipation near $g_c$. The effect of dissipation is to prepare the \textit{particular} steady state in the Hilbert space over time, which leads to damping of the peaks over time. We notice early time oscillations because  coherent dynamics dominate over dissipation; however, for longer times, $L(t)$ no longer exhibits sustained revivals, but instead shows a rapid initial decay followed by relaxation towards a steady value. It saturates to a nonzero plateau that depends on both $N$ and $\gamma/\Omega$. Stronger dissipation (dashed curves) suppresses oscillations more efficiently and drives the system faster to the steady regime, whereas weaker dissipation (solid curves) allows transient coherence to persist for longer times. For large $N$, the quantity becomes smoother and is less dependent on time.

Fig. \ref{fig6}(c) remarks the dissipative dynamics for the time evolution of Loschmidt echo in the presence of Rydberg interactions. We note that the dynamics become strongly interacting and effectively irreversible. The Loschmidt echo grows monotonically toward a large steady value, rather than exhibiting oscillatory relaxation. Interactions largely redistribute quantum information across many-body configurations, while dissipation removes coherence, producing a smooth approach to equilibrium that is sensitive to system size $N$. A larger $N$ leads to slower but higher saturation values $L_s$. 

\begin{align}
    L(t) \to L_s.
\end{align}
We note that the Loschmidt echo tracks the transition from coherent many-body revivals to environmentally induced irreversibility under the combined effects of interactions and dissipation.

\subsection{Von Neumann Entropy}
We determine the entropy of the spin and phonon degree of freedom in the ion string which is calculated from the reduced density matrix as,
\begin{align}
    S_{\text{spin}} & = -\text{Tr}(\rho_{\text{spin}}\ln\rho_{\text{spin}}) ,\nonumber \\
    S_{\text{ph}} & = -\text{Tr}(\rho_{\text{ph}}\ln\rho_{\text{ph}}),
    \label{entropy}
\end{align}
where $\rho_{\text{spin/ph}}=\text{Tr}_{\text{ph/spin}}[\rho(t)]$ is the partial trace performed over the phonon or spin degree of freedom respectively~\cite{amicoEntanglementManybodySystems2008}. It measures the degree of purity or mixedness in the reduced spin or phonon state respectively due to the interaction process and environmental coupling.
The maximum possible entropy of a subsystem with Hilbert space dimension `$d$' is $S_{\max} = \ln d$
when the subsystem is maximally mixed~\cite{PhysRevA.67.022110}.

\subsubsection{Spin Entropy}
Fig. \ref{fig7}(a)-(c) show the time evolution of normalized spin entropy $S_{\text{spin}}$ of the ion subsystem with varying $N$ for different cases.  In the absence of dissipation, one observes that the entropy increases with finite pronounced oscillations and partial revivals. It happens because the spin phonon coupling continuously redistributes quantum amplitudes between the spin and phonon sectors, causing the reduced internal spin state to evolve from an initially pure state into a progressively more mixed state (Fig. \ref{fig7}(a)). Moreover, the spin entropy reflects the reversible dynamics due to coherent interactions. The reduced spin state becomes increasingly mixed, leading to higher entropy for larger system sizes.

In the presence of dissipation, the entropy saturates to a nearly constant value in short evolution times (Fig. \ref{fig7}(b)) as it drives the system towards a mixed steady state. Stronger dissipation ($\gamma/\Omega=1.5$) suppresses coherent population redistribution more efficiently and drives the spins more rapidly toward the steady state, yielding lower entropy values.
Increasing $N$ further reduces the final entropy as dissipation effects are larger with increasing $N$ and stronger dissipation drives the spins more rapidly toward the steady state, yielding lower entropy values. 

In Fig. \ref{fig7}(c), with both dissipation and Rydberg interactions present, the entropy production becomes strongly interaction-assisted. An initially pure spin state rapidly develops finite entropy, indicating that the ion state becomes mixed due to its interaction with both the phonon mode and the dissipative environment. The ionic entropy $S_{\text{ion}}$ exhibits two distinct dynamical regimes: an initial rapid exponential increase arising from fast interaction-induced correlation buildup and dissipative mixing, followed by a slower linear growth regime associated with collective relaxation towards the steady state with lower entropy value with increasing $N$. Stronger dissipation accelerates the relaxation of the ionic state towards the steady state, leading to a more rapid transition to a lower entropic state.

\subsubsection{Phonon Entropy}
\begin{figure}[t]%
\includegraphics[width=\linewidth]{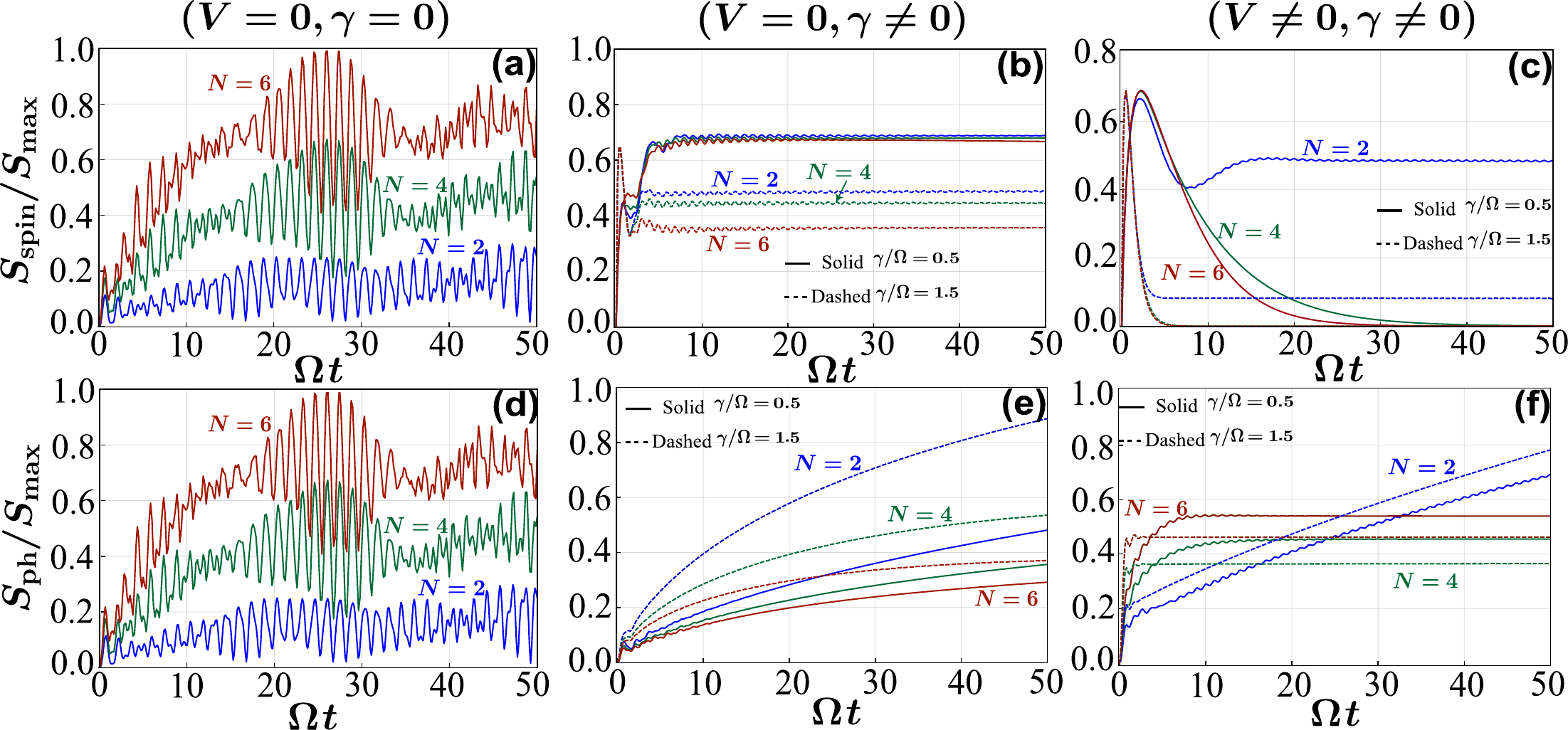}
\caption{Time variation of Von Neumann Entropy of (a)-(c) the spins and (d)-(f) the phonon mode~\eqref{entropy} with different system sizes $N$ for various cases: Closed Dicke dynamics ($V=0,\gamma=0$); (b) dissipative Dicke dynamics ($V=0,\gamma \neq 0$); (c) dissipative dynamics with Rydberg interactions ($V \neq 0,\gamma \neq 0$).  The solid and dashed lines denote the dynamics for dissipation strengths $\gamma/\Omega=0.5$ and $\gamma/\Omega=1.5$ respectively. We choose $g/\Omega=0.5$  for all cases and $\mathcal{V}/\Omega=5$.  
}
\label{fig7}%
\end{figure}

Fig. \ref{fig7}(d)-(e) shows the time evolution of the normalized phonon entropy with varying $N$. In the absence of dissipation, the combined spin-phonon system evolves unitarily and therefore remains in a globally pure state. Consequently, the entropy of the reduced phonon subsystem is identical to that of the reduced ionic subsystem at all times as shown in Fig. \ref{fig7}(d). One can observe partial revivals because of the coherent exchange of excitations between spins and phonons. 

In the presence of excited state decay, the phonon entropy initially increases monotonically and then approach a growing mixed state (Fig. \ref{fig7}(e)). Stronger dissipation and a large $N$ suppresses the buildup of mixedness generated by the coherent dynamics and stabilizes the phonon state more efficiently. One major difference is that, unlike spin entropy, phonon entropy has no upper bound because of the infinite Hilbert space associated with phonons. Because the phonon space is unbounded, the entropy does not saturate to a fixed maximum value. Moreover, the phonon entropy value is larger with increasing dissipation parameter as the energy is transferred from the spins to the environment via the phonon mode.

The presence of Rydberg interactions (Fig. \ref{fig7}(f)) allows the phonon entropy to remain comparatively constant after a short transient compared to the cases shown in Fig. \ref{fig7}(d),(e). It strongly suppresses the accumulation of mixedness in the phonon subsystem. Therefore, the combined action of Rydberg interactions and dissipation produces the most robust dynamical regime, where coherent spin-phonon mixing is efficiently suppressed and the phonon subsystem remains comparatively pure throughout the long time evolution.

\section{Conclusion}
\label{sec5}
In this work, we have investigated the non equilibrium dissipative dynamics of a generalized Dicke model with a system of trapped Rydberg ions, where the system dynamics is subjected to collective spin-phonon coupling, atomic dissipation defined by Markovian process, and Rydberg-mediated ion-ion interactions generating a rich dynamical behavior. We employ a MF approach to analyze the steady-state solutions and construct the dissipative phase diagram. Our analysis reveals the existence of bright and dark phases, coexistence regions, and interaction-induced dynamical instabilities and tricritical point. In particular, we show that the inclusion of Rydberg interactions introduces an additional nonlinear interaction channel that fundamentally modifies the phase structure of the conventional dissipative Dicke model, giving rise to interaction-induced coexistence regimes and oscillatory dynamics associated with the Hopf bifurcations.

To further characterize the system beyond MF approximation, we perform exact finite-size simulations by solving the Lindblad master equation. The time evolution of collective spin observables, ion-state populations, von Neumann entropy, and the Liouvillian spectrum is analyzed for different system sizes. These calculations allow us to assess the validity of the MF predictions and investigate the role of finite-size effects and quantum fluctuations. We find that the finite-size dynamics 
exhibit signatures of slow relaxation, metastability, and size-dependent behavior near the phase boundaries. Furthermore, the analysis of the Liouvillian spectrum indicates a progressive closing of the Liouvillian gap with increasing system size, indicating the emergence of long-lived dynamical modes and the onset of dissipative critical behavior. It is apparent that systems show oscillatory dynamics, dissipative relaxation through long-lived transient dynamics associated with coexistence phases and finally to interaction-stabilized steady states.

We also explore the non equilibrium response of the system under sudden quenches of the control parameters. By introducing the Loschmidt echo as a diagnostic tool, we characterize the dynamical phase transitions and quantify the sensitivity of the time-evolving quantum state to abrupt parameter changes. In the absence of dissipation, the Loschmidt echo exhibits pronounced revivals associated with coherent spin-phonon dynamics, whereas dissipation progressively suppresses these revivals and drives the system toward steady-state behavior. We infer that the interplay of Rydberg ion interactions and dissipation strongly influence the relaxation pathways and dynamical transitions between different phases.

Finally, our results demonstrate that trapped Rydberg-ion systems provide a versatile and experimentally accessible platform for studying dissipative many-body physics and dynamical phase transitions. Our model is formulated within the Markovian approximation and demonstrates the existence of tricritical points associated with multiple nonequilibrium phases arising from the interplay of spin-phonon coupling, dissipation, and Rydberg interactions. We invoke finite-size effects to understand dissipative critical phenomena and establish the Loschmidt echo, entropy growth, and Liouvillian spectral properties as powerful probes of dynamical phase transitions in open quantum systems. These findings pave the way for future investigations of nonequilibrium quantum criticality, metastability, and dissipative-state engineering in trapped-ion and other hybrid quantum platforms. The present framework provides a foundation for several future directions, including the incorporation of multimode phonon dynamics and non-Markovian regime, which can further control the nonequilibrium phase structure and quantum dynamics of trapped Rydberg ion systems.

\section*{Acknowledgements}


\paragraph{Funding information}
W.L. acknowledges support from the EPSRC through Grant No. EP/W015641/1, and the Going Global Partnerships Programme of the British Council (Contract No. IND/CONT/G/22-23/26).

\bibliography{SciPost_Example_BiBTeX_File.bib}


\end{document}